\documentclass[prd,a4paper,nofootinbib,doblespace
]{revtex4}
\usepackage{graphicx}
\usepackage{amsmath}
\usepackage{multirow}
\newcommand{\nn}{\nonumber}

\newcommand{\beq}{\begin{equation}}
\newcommand{\eeq}{\end{equation}}
\newcommand{\bea}{\begin{eqnarray}}
\newcommand{\eea}{\end{eqnarray}}
\newcommand{\ba}{\begin{array}}
\newcommand{\ea}{\end{array}}
\newcommand{\bec}{\begin{center}}
\newcommand{\eec}{\end{center}}
\newcommand{\bei}{\begin{itemize}}
\newcommand{\eei}{\end{itemize}}

\usepackage{pstricks}
\usepackage{graphicx}
\usepackage{inputenc}
\usepackage{pst-coil}
\usepackage{longtable}
\usepackage{amssymb}
\usepackage{amsfonts}
\usepackage{bbm}
\usepackage{color}

\def\eq#1{{eq.~(\ref{#1})}}

\def\vev#1{\left\langle #1\right\rangle}

\def\hbar{\hspace{0pt}\raisebox{1pt}{$-$} \hspace{-7pt} h}

\def\5{\bar 5}

\begin{document}

\vspace{0.3cm}

\title{Tri-bi-maximal mixing in viable family symmetry unified model with extended seesaw}
\author{
Federica Bazzocchi$^1$ and Ivo de Medeiros Varzielas$^2$\\
$^1$~Department of Physics and Astronomy, Vrije Universiteit Amsterdam,\\
1081 HV Amsterdam, The Netherlands\\
$^2$ CFTP, Departamento de F\'{i}sica, Instituto Superior T\'{e}cnico,\\
Av. Rovisco Pais, 1, 1049-001 Lisboa, Portugal
}

\begin{abstract}
We  present a Grand Unified model based on $SO(10)$ with a $\Delta(27)$ family symmetry. Fermion masses and mixings are fitted and agree well with experimental values. An extended seesaw mechanism plays a key role in the generation of the leptonic mixing, which is approximately tri-bi-maximal. 
\end{abstract}

\maketitle

\section{Introduction}

The observed masses and mixings of the fermions, and the existence of three families of fermions are left unexplained by the Standard Model (SM). This is just one of many theoretical motivations for going beyond the SM, either through Grand Unified Theory (GUT) extensions or by adding a Family Symmetry (FS), usually using Supersymmetry (SUSY) to keep the hierarchy problem under control.
Adding a FS to justify the patterns of fermion parameters is motivated by the observation of leptonic mixing consistent with, and in fact well approximated by Tri-Bi-Maximal (TBM) mixing \cite{Wolfenstein:1978uw, Harrison:2002er, Harrison:2002kp, Harrison:2003aw, Low:2003dz}.

There are many FS models in the literature that obtain mixing that is close to TBM for the leptons, but there are relatively few that simultaneously justify the large leptonic mixing and the strong hierarchy and small mixing of the quark sector. Ambitious FS models that tackle both the lepton and quark sectors usually do so using the Froggatt-Nielsen (FN) mechanism \cite{fnielsen} in the context of a SUSY GUT symmetry commuting with the FS \cite{King:2005bj, Ivo1, Ivo3, Morisi:2007ft, Bazzocchi:2008rz}.
In particular, considering an underlying $SO(10) \times G_F$ structure is highly constraining as the Left-Handed (LH) and Right-Handed (RH) SM fermions must then transform the same under $G_F$, to be unified consistently into a single multiplet (the $16$ of $SO(10)$).

Implementing the seesaw mechanism \cite{Minkowski:1977sc, Yanagida:1979as, Glashow:1979nm, Mohapatra:1979ia, GellMann:1980vs, Schechter:1980gr} in a non-minimal way \cite{Mohapatra:1986aw, Mohapatra:1986bd, Barr:2003nn, Malinsky:2005bi} requires an enlarged field content. GUT FS models use multiple familon fields to break the FS, so requiring the neutrino sector to be minimal without considering the context is not readily justified - for example, the added freedom in the neutrino sector may enable a reduction in the number of familons needed. It is thus interesting to consider the possible benefits that can be derived from combining a FS with extended seesaw. Recently, in \cite{Hagedorn:2008bc}, those two ingredients are used to provide an explicit realization of the screening mechanism \cite{Lindner:2005pk}.

The SUSY GUT model we present relies on the (extended) seesaw mechanism and on a discrete FS to obtain TBM mixing. Subtly, the non-minimal structure of the neutrino sector allows some freedom in choosing the field content of the model (e.g. our model does not include any $45$ representations, which are ubiquitous in other $SO(10) \times G_F$ models \cite{Ivo1, Ivo3, Bazzocchi:2008rz}).

\section{The model \label{mod}}

The superfields and their representations under the symmetry content of the model are summarized in table \ref{SO10}. We start with an underlying $SU(3)_{F}$ FS as in practice the effects of considering its discrete subgroup, $\Delta(27)$, are relevant for the VEV alignment only (see section \ref{f_align}).

\begin{table}[t]
\begin{center}
\begin{tabular}{|c|c|c|c|c|c|c|}
\hline
\multicolumn{7}{|c|}{Matter fields} \\ \hline
\multicolumn{2}{|c|}{} & $\Psi$ & $\Psi_\eta$ & $\bar{\Psi}_\eta$ & $s$&$\bar{s}$ \\ \hline
\multicolumn{2}{|c|}{$SO(10)$} & 16 & 16 & $\bar{16}$ & 1 & 1 \\ 
\multicolumn{2}{|c|}{$SU(3)_{F}$} & 3 & 1 & 1 & 3 & $\bar{3}$ \\ 
\multicolumn{2}{|c|}{$U(1)_F$} & 1 & $q_{\eta}$ & $-q_{\eta}$ & $q_{s}$ & $-q_{s}$ \\ \hline \hline
\multicolumn{7}{|c|}{Familons} \\ \hline
\multicolumn{2}{|c|}{} & $\phi_{23}$ & $\phi_{123}$ & $\phi_3$ & $\phi_O$ & $\phi_0$ \\ \hline
\multicolumn{2}{|c|}{$SO(10)$} & 1 & 1 & 1 & 1 & 1 \\
\multicolumn{2}{|c|}{$SU(3)_{F}$} & $\bar{3}$ & $\bar{3}$ & $\bar{3}$& $\bar{3}$ & 1 \\
\multicolumn{2}{|c|}{$U(1)_F$} & $q_{23}$ & $q_{123}$ & $-3 q_{s} -2 q_{123}$ & $q_{23}-q_{s}-q_{123}$ & $2q_{s}+2 q_{123}$ \\ \hline \hline
\multicolumn{7}{|c|}{Higgs fields} \\ \hline
 & $\varphi$ & $\varphi'$ &$\tilde{\varphi}$& $\xi$ & $\rho$ & $\Sigma$ \\ \hline
$SO(10)$ & $10$  & $10$ &$10$& $\bar{16}$ & $\bar{126}$ & $210$ \\
$SU(3)_{F}$ & 1 & 1 & 1 & 1 & 1 & 1 \\
$U(1)_F$ & $-2 q_{\eta}$ & $3 q_{s}+2 q_{123}-q_{\eta}$ &$2q_{s}+ 2q_{123}-2 q_{23}$& $-q_{s}-2 q_{123} $ & $-2 q_{23}$ & $3 q_{s}+2 q_{123}+q_{\eta}$ \\ \hline
\end{tabular}
\end{center}
\caption{Chiral superfields and their charges. The $U(1)_{F}$ charges of $10$s of $SO(10)$ and of the $\bar{3}$s of $SU(3)_{F}$ must be unique (e.g. $q_{23} \neq q_{123}$).}
\label{SO10}
\end{table}
The $U(1)_{F}$ charge of $\phi_O$ is specified such that $\phi_{23} \phi_{23}$ has the same overall charge assignment as $\phi_{O} \phi_{O} \phi_{0}$.

The singlets $s$ and $\bar{s}$ enlarge the neutrino sector leading to extended seesaw.
$\Psi$ contains the SM fermions and the RH neutrinos.
The $\eta$ fields $\Psi_{\eta}$ and $\bar{\Psi}_{\eta}$ serve as  FN-like  messenger fields and behave as a fourth heavy family that mixes with the third family of the SM fermions.

The Higgs sector breaks $SO(10)$ down to the SM gauge group.
The Vacuum Expectation Value (VEV) of a Higgs field is generically denoted as $v^F_{A}$, the label $F$ denoting the field and the label $A$ denoting the $SO(10)$ breaking direction with respect to $SU(5)$. For example, $v^{\Sigma}_{75}$. In this notation it is useful to keep in mind the $SU(5)$ representations within the $SO(10)$ representations that contain them ($\Sigma$ is a $210$ of $SO(10)$, containing a $75$ with respect to $SU(5)$).

The familons are $SO(10)$ singlets and break the FS when they acquire a non-vanishing VEV, and are generically denoted as $\phi_A$ ($\vev{\phi_A}$ for the corresponding VEV). The label in $\phi_0$ denotes the field is an $SU(3)_F$ singlet (note it is however charged under $U(1)_F$). The labels $A = 3, 23, 123$ serve to identify the direction of the respective VEVs (the numbers identify which entries do not vanish), and the label in $\phi_{O}$ denotes ``Orthogonal'' (its VEV is orthogonal to both the $23$ and $123$ VEVs). Specifically, the VEVs direction are given by
\begin{eqnarray}
\label{f_vevs}
\vev{\phi_{3}} \propto \left( 0,0,1 \right) , \nn\\
\vev{\phi_{23}} \propto \left( 0,1,-1 \right) , \nn\\
\vev{\phi_{123}} \propto \left( 1,1,1 \right) , \nn\\
\vev{\phi_{O}} \propto \left( 2,-1,-1 \right) .
\end{eqnarray}

While the Lagrangian must be invariant under $\Delta(27)$, in practice the terms allowed by the discrete FS (and not by $SU(3)_F$) require distinct messengers, and are either absent or present only at higher order such that they are strongly suppressed (the $\Delta(27)$ invariants in the real potential are also very small, but as the only terms that distinguish the VEV directions they can not be neglected in the alignment discussion). The Lagrangian invariant under the symmetry content in Table \ref{SO10} is given by
\begin{eqnarray}
\label{lag}
\mathcal{L}_Y&=&
\frac{1}{\Lambda^2} (\phi_{23} \Psi )\, \rho\,(\phi_{23} \Psi ) +
\frac{1}{\Lambda^2}(\phi_{123} \Psi )\,\xi\, (\phi_{123} s) +
\frac{1}{\Lambda^3}(\phi_O \Psi)\,\rho\, (\phi_O \Psi) \phi_0 +
M_s\, (\bar{s} s) +
\frac{1}{\Lambda}  (\bar{s} \Psi) \xi \phi_0 \nn\\
& +&
\frac{1}{\Lambda}(\phi_3 \Psi) \,\Sigma\,\bar{\Psi}_\eta +
M_\eta \,\bar{\Psi}_\eta\, \Psi_\eta +
\frac{1}{\Lambda^2}(\phi_O \Psi ) \tilde{\varphi}\, (\phi_O \Psi) +
\frac{1}{\Lambda^3} (\phi_3 s) (\phi_3 s) \phi_0^2 +
\frac{1}{\Lambda} (\phi_3 \Psi ) \varphi' \,\Psi_\eta+\Psi_\eta \, \varphi \, \Psi_\eta .
\end{eqnarray}
The $U(1)_F$ assignments ensure that any undesirable terms are absent or sufficiently suppressed.
Parentheses denote the $SU(3)_{F}$ invariant contractions $(\phi_{23} \Psi) = \phi_{23}^{i} \Psi_{i}$, with other contractions not allowed or suppressed by the messenger content (e.g. $(3\otimes 3) \otimes (\bar{3} \otimes \bar{3})$ contractions are absent). The non-renormalizable terms have associated the cut-off scale $\Lambda$, assumed to be $\Lambda \sim 10^{17} \,\mbox{GeV}\,> M_{GUT}\sim 2 \times 10^{16}\,\mbox{GeV}$. The other mass scales are the $\eta$ messengers mass and the singlet masses, $M_\eta \sim M_s\sim 10^{12}-10^{14}$  GeV.

\subsection{Neutrino masses and extended seesaw \label{neutrinos}}

To obtain viable leptonic mixing, we aim to generalize the method described in detail in \cite{Ivo6} to extended seesaw mechanisms. Before proceeding with this generalization, we use component notation explicitly in order to illustrate how one can achieve TBM neutrino mixing through Type I seesaw.
In Type I seesaw the effective neutrino mass matrix is given in component notation by
\begin{equation}
\label{comp_seesaw}
(m_{\nu})^{ij} = - (m_{D})^{il} \, (M_{R}^{-1})_{lk} \, (m_{D}^{T})^{kj} .
\end{equation}
$m_{D}$ is the neutrino Dirac matrix and $M_{R}$ is the heavy RH Majorana neutrino mass matrix.

In FS models the mass matrices are typically given by some combination of the familon VEVs. Specifically in the type of model considered here the Dirac mass can be written as
\begin{equation}
\label{comp_Dirac}
(m_{D})^{il}= \vev{\phi_{A}^i}^{T} \,\vev{\phi_{C}^l} .
\end{equation}
We have omitted any proportionality constants that have no family index structure. The components of $m_{D}$ are clearly given by the familon VEVs family structure.
Inserting $m_{D}^{il}$ into eq.(\ref{comp_seesaw}) we have:
\begin{equation}
\label{comp_eff}
(m_{\nu})^{ij} = - \vev{\phi_{A}^i}^{T} \, \left( \vev{\phi_{C}^l} \,  (M_{R}^{-1})_{lk}\, \vev{\phi_{C}^k}^{T} \right)  \vev{\phi_{A}^j} .
\end{equation}
Note that the quantity $a =  \vev{\phi_C^l} \,  (M_{R}^{-1})_{lk}\, \vev{\phi_C^k}^{T} $ is just a constant with no index structure, and therefore:
\begin{equation}
\label{str1}
(m^{\nu})^{ij} = - a\vev{\phi_A^i}^{T} \,  \vev{\phi_A^j} .
 \end{equation}
Unless $a=0$, $\vev{\phi_A}$ is an eigenstate of $m^{\nu}$ and the details of $\vev{\phi_{C}}$ and $M_R$ only serve to determine the corresponding eigenvalue.

Generalizing, with:
\begin{equation}
\label{str2}
(m_D)^{il} = a' \vev{\phi_{A}^i}^{T} \,\vev{\phi_{C}^l} \, + b' \vev{\phi_{B}^i}^{T} \,\vev{\phi_{D}^l} .
\end{equation}
we have the corresponding effective neutrino matrix:
\begin{equation}
\label{comp_eff_general}
(m_{\nu})^{ij} = - [a \vev{\phi_A^i}^{T} \,  \vev{\phi_A^j} \, + b \vev{\phi_B^i}^{T} \,  \vev{\phi_B^j} \,
+ c \vev{\phi_A^i}^{T} \,  \vev{\phi_B^j} + d \vev{\phi_B^i}^{T} \,  \vev{\phi_A^j} \,] ,
\end{equation}
where $a, b$ and $c, d$ are constants involving the products of the respective VEVs $\vev{\phi_{C}}$, $\vev{\phi_{D}}$ with $M_{R}^{-1}$.
From eq.(\ref{comp_eff_general}) we conclude that as long as $\vev{\phi_{A}}$ and $\vev{\phi_{B}}$ are orthogonal, they are both eigenstates of $m_{\nu}$ provided that $c = d = 0$. The natural expectation in GUT FS models is that $M_{R}$ is structured similarly to $m_{D}$ (in terms of being analogously formed by familon VEVs), and then one can identify which combinations of familons in $M_{R}$ leads to $c = d = 0$.

After establishing how the method works for Type I seesaw, it is straightforward to apply it to extended seesaw: if the extended seesaw gives as a result eq.(\ref{comp_eff_general}) with generalised $a, b$ and $c = d = 0$ numbers we can still easily identify the eigenvectors. The difference is that instead of the Type I relation (e.g. $a = \vev{\phi_C^l} \,  (M_{R}^{-1})_{lk}\, \vev{\phi_C^k}^{T} $) these numbers will be in general more complicated products of the respective intervening familon VEVs and the relevant neutrino matrices of extended seesaw.
Although the following details may be somewhat complicated due to the intricacies of the GUT and of the extended seesaw, the basic idea is rather simple - we want to obtain TBM mixing in the neutrino sector directly from two orthogonal VEVs, $\vev{\phi_{123}}$ and $\vev{\phi_{23}}$. In the particular realization we consider, the tri-maximal eigenstate is obtained through a linear seesaw that starts from the term $(\phi_{123} \Psi) (\phi_{123} s)$ (this eigenstate has to arise through extended seesaw as the starting term involves the singlet $s$). In contrast, the bi-maximal eigenstate is obtained through both Type I and Type II seesaw resulting from $(\phi_{23} \Psi) (\phi_{23} \Psi)$. In this particular realization we also produce the orthogonal eigenstate explicitly from $(\phi_{O} \Psi) (\phi_{O} \Psi)$ (similarly to the bi-maximal state).

In order to consider in detail how the seesaw proceeds, we write the full neutrino mass matrix $M_{\nu}$. We do not need to consider $\Psi_{\eta}$ mixing in the neutrino sector (the $\eta$ mixing is considered in detail in section \ref{charged}) due to the VEVs of the Higgs sector - particularly, $\vev{\Sigma}$ develops only along the $75$ of $SU(5)$. We start in the $(\nu,\nu^c, s,\bar{s})$ basis (each of these fields is a triplet under the FS). It is convenient to write the $12 \times 12$ $M_{\nu}$ as a $4 \times 4$ block matrix (each block is $3 \times 3$)
\begin{eqnarray}
\label{eq:matneut0}
M_{\nu} = 
\left( \begin{array}{cccc}
m_{LL}& . & . & . \\
m_{R L} & m_{RR} & . & .\\
m_{S L} & m_{ S R} & m_{ SS} & .\\
m_{ \bar{S} L} & m_{ \bar{S} R}&m_{ \bar{S} S} &0
\end{array}\right) ,
\end{eqnarray}
with
\begin{eqnarray}
\label{blockexp}
m_{LL} &=& v^{\rho}_{15}( \vev{\phi_{23}}^{T} \vev{\phi_{23}}/\Lambda^2+ \vev{\phi_{O}}^{T} \vev{\phi_{O}} \vev{\phi_0}/\Lambda^3) , \nn\\
m_{RL} &=&  v^{\rho}_{5} (\vev{\phi_{23}}^{T} \vev{\phi_{23}}/\Lambda^2 + \vev{\phi_{O}}^{T} \vev{\phi_{O}}\vev{\phi_0}/\Lambda^3 ) , \nn\\
m_{RR} &=&   v^{\rho}_{1} ( \vev{\phi_{23}}^{T} \vev{\phi_{23}}/\Lambda^2 + \vev{\phi_{O}}^{T} \vev{\phi_{O}} \vev{\phi_0}/\Lambda^3) , \nn\\
m_{SL}&=& v^{\xi}_{5} \vev{\phi_{123}}^{T} \vev{\phi_{123}}/\Lambda^2 , \nn\\
m_{S R} &=&  v^{\xi}_{1} \vev{\phi_{123}}^{T} \vev{\phi_{123}}/\Lambda^2 , \nn\\
m_{S S} &=&  \vev{\phi_0}^{2} \vev{\phi_{3}}^{T} \vev{\phi_{3}}/\Lambda^3 , \nn\\
m_{\bar{S} L} &=&  v^{\xi}_{5} I \vev{\phi_0}/\Lambda , \nn\\
m_{\bar{S} R} &=&   v^{\xi}_{1} I \vev{\phi_0}/\Lambda , \nn\\
m_{\bar{S} S} &=&  M_s  \, I .
\end{eqnarray}
$I$ is the $3 \times 3$ identity matrix. Note that $m_{SS}  \propto Diag (0,0,1)$. $M_{\nu}$ is symmetric and we use dots in redundant blocks. The Higgs VEVs follow the notation discussed in the introduction, with the subscript labels corresponding to the VEV that projects the appropriate components of the matter fields in each block (e.g. $v^{\rho}_{1}$ and $v^{\xi}_{1}$ appear in the RH neutrino blocks, correspond to $SU(5)$ singlets, and project the RH neutrino component of $\Psi$, $\nu^c$).

We assume that $m_{\bar{S} S} > m_{RR}  > m_{SS}, m_{S R},m_{\bar{S} R} $. We first consider the $9 \times 9$ sub-block that leaves out the first three ($\nu$) rows and columns and go into the basis in which $\bar{s}$ and $s$ form a Dirac spinor. Continuing to use the $3\times 3$ blocks defined above
\begin{eqnarray}
\label{eq:matneut1}
\left( \begin{array}{ccc}
m_{RR} & . & . \\
m_{SR} &m_{SS} & . \\
m_{\bar{S} R} & m_{\bar{S} S}& 0
\end{array}\right) ,
\end{eqnarray}
becomes approximately
\begin{eqnarray}
\label{eq:matneut2}
\left( \begin{array}{ccc}
M_{RR} & . & .\\
0 & M_s I + m_{SS}/2 & .\\
0 & m_{SS}/2 &-M_s I + m_{SS}/2
\end{array}\right) ,
\end{eqnarray}
with 
\begin{equation*}
M_{RR} = m_{RR}+\tilde{M} ,
\end{equation*}
\begin{equation}
\tilde{M} = \frac{2}{M_s}\,m_{SR}^T m_{\bar{S} R}  + \frac{2}{M_s^2}[m_{\bar{S} R}^T m_{SS}  m_{\bar{S} R}] .
\end{equation}
Note that the combination inside square brackets, while seemingly complicated, is simply proportional to $m_{SS}$.

Consistently re-introducing the $\nu$ part of $M_{\nu}$, we have in the new basis
\begin{eqnarray}
\label{eq:matneut3}
M_{\nu} \simeq \left( \begin{array}{cccc}
m_{LL} & . & . & . \\
m_{RL} & M_{RR} & . & .\\
(m_{SL}+ m_{\bar{S} L}) /\sqrt{2} & 0 & M_s I + m_{SS}/2 & .\\
(m_{SL}- m_{\bar{S} L}) /\sqrt{2} & 0 & m_{SS}/2 &-M_s I + m_{SS}/2
\end{array}\right) ,
\end{eqnarray}
from where we can read off the $3 \times 3$ light Majorana neutrino mass matrix structure $m_{\nu}$:

\begin{equation}
\label{massnu}
m_\nu= m_{LL}+ m_{RL}^T M_{RR}^{-1} m_{RL} + \frac{2}{M_s}  m_{SL}^T  m_{\bar{S} L} + \frac{2}{M_s^2} [ m_{\bar{S}L}^T m_{SS} m_{\bar{S}L}] .
\end{equation}

The third term  is the linear seesaw contribution arising by the extra singlets and produces the candidate tri-maximal eigenstate rather trivially, as its structure is
\begin{equation}
\vev{\phi_{123}}^T \vev{\phi_{123}} ,
\end{equation}
as the other matrices involved in the term are proportional to $I$.

The first  term  in \eq{massnu} is the type II seesaw contribution, of the form
\begin{equation}
a'' \vev{\phi_{23}}^T \vev{\phi_{23}} +  b'' \vev{\phi_O}^T \vev{\phi_O} ,
\end{equation}
The  Dirac mass matrix that enters in the second term of \eq{massnu} presents the general structure given in eq.(\ref{str1}) since   from \eq{blockexp} we have (similarly to eq.(\ref{str2})
\begin{equation}
m_{RL} = a' \vev{\phi_{23}}^{T} \vev{\phi_{23}}+ b' \vev{\phi_{O}}^{T} \vev{\phi_{O}} ) .
\end{equation}
The orthogonality between $\vev{\phi_{23}}$ and $\vev{\phi_{O}}$ ensures that the coefficients equivalent to the $c,d$ of \eq{comp_eff_general} vanish and therefore the contribution to $m_\nu$ arising by the second term presents the same structure as $m_{LL}$
\begin{equation}
a \vev{\phi_{23}}^T \vev{\phi_{23}}+   b \vev{\phi_O}^T \vev{\phi_O} .
\end{equation}
The resulting effect is that we obtain a candidate bi-maximal eigenstate, and also explicitly a candidate third eigenstate in TBM mixing (orthogonal to both the tri-maximal and bi-maximal eigenstates).

The last term in eq.(\ref{massnu}) is  proportional to $Diag(0,0,1)$, incompatible with TBM mixing. Fortunately it is rather suppressed through what might be thought of as a generalisation of Sequential Dominance \cite{King:1998jw, King:1999cm, King:1999mb, King:2002nf, Ivo6} - the resulting magnitude is approximately $10^{-2} \sqrt{\Delta m_{sol}^2}$, therefore we can neglect it to good approximation. The $\phi_{O}$ candidate orthogonal eigenstate is also suppressed due to $\vev{\phi_{0}}$. The $\phi_{123}$ and $\phi_{23}$ states are naturally heavier in this scheme so a Normal Hierarchy is predicted for the effective neutrinos.

To summarize, we concluded that the effective neutrino mass matrix is given by
\begin{eqnarray}
m_\nu&\simeq& \alpha \vev{\phi_{23}}^T \vev{\phi_{23}}+  \beta \vev{\phi_{123}}^T \vev{\phi_{123}}+ \gamma \vev{\phi_O}^T \vev{\phi_O}\,\nn\\
&=&\left( \begin{array}{ccc}  \beta+4 \gamma & \beta -2 \gamma &\beta -2\gamma \\ \beta -2 \gamma & \alpha +\beta +\gamma &-\alpha+\beta +\gamma \\ \beta -2 \gamma  &-\alpha+\beta +\gamma & \alpha +\beta+\gamma \end{array}\right) ,
\end{eqnarray}
where for clarity we absorbed the magnitude of the VEVs such that $\vev{\phi_{23}}$, $\vev{\phi_{123}}$, $\vev{\phi_{O}}$ have integer entries (appropriately defining $\alpha$, $\beta$, $\gamma$). This form satisfies $(m_\nu)_{11}= (m_\nu)_{22}+(m_\nu)_{23}-(m_\nu)_{13}$ and it is diagonalized by TBM mixing, becoming $m_\nu^{diag}=Diag(6 \gamma, 3 \beta, 2\alpha)$ with eigenvalues $6 \gamma \ll 3 \beta < 2 \alpha$.

In order to fit the neutrino mass splitting data we need $v^{\xi}_1\sim 10^{10} - 10^{12}$ GeV, $v^{\rho}_1 \sim 10^{12}-10^{14}$ GeV and $M_s\sim 10^{14}$ GeV, for familon VEVs satisfying  $\vev{\phi_0}/\Lambda\sim \lambda^3$ and $\vev{\phi_{23}^{2,3}}/\Lambda \sim \lambda$, $\vev{\phi_{3}^{3}}/\Lambda \sim \sqrt{\lambda}$ where $\lambda$ is the Cabibbo angle. As we shall see in the next sections these values for the familon VEVs are fixed once we take into account the  charged fermion spectrum. The magnitude of $\vev{\phi_{123}^{i}}$ is not tightly constrained by phenomenology as its VEV is associated always with $v^\xi_{5}$. For alignment purposes we take $\vev{\phi_{123}^{i}}$ to be large compared to the other familon VEVs (see section \ref{f_align}). 

\subsection{Charged fermion masses \label{charged}}

We will now describe how the charged fermion mass hierarchies are obtained and how the quark mixing is generated. The SM  fermions belong to the $16$s of $SO(10)$. We denote the $10$ of $SU(5)$ inside $\Psi$ and $\bar{\Psi}_{\eta}$, as $f_i$ and $\bar{f}_\eta$ respectively (this notation separates e.g. the lepton doublets $L_{i}$).
When $\phi_3$ and $\Sigma$ develop their VEVs  ($\vev{\Sigma}=v^\Sigma_{75}$), the term $(\phi_3 \Psi) \,\Sigma\,\bar{\Psi}_\eta$ in the Lagrangian (\eq{lag}) becomes

\begin{equation}
\label{quark}
 \bar{f}_\eta (\alpha_f  v_3 v^\Sigma_{75}\,f_{3} + M_\eta f_\eta)\,, \end{equation}
defining $v_3=\vev{\phi_{3}^{3}}/\Lambda$ (the non-zero VEV of the $i=3$ component of $\phi_3^i$).
$\alpha_f$ is a Clebsch-Gordan factor and we assume $v_{75}^\Sigma,\vev{\phi_{3}^{3}}  \sim 10^{16}\,\mbox{GeV}\, \gg M_\eta$.
Due to $\Sigma$, the mixing with the $\eta$ field involves only $Q_i,u^c_i,e^c_i$, not involving $d^{c}_i$, and also not involving $L_i$ which keeps the $\eta$-mixing from strongly affecting the leptonic mixing angles.
We define the heavy and light  combinations
\begin{eqnarray}
\label{basisquark}
f_h &=&  s_f  f_\eta+c_f \, f_{3} , \nn\\
f_{l_3} &=&  -s_f f_{3}+c_f \,f_{\eta} ,
\end{eqnarray}
with
\begin{equation}
\label{defang}
s_f= \frac{M_\eta}{\sqrt{\alpha_f^2 v_3^2  (v^\Sigma_{75})^2+M_\eta^2}} \quad , \quad c_f= \frac{\alpha_f v_3 v^\Sigma_{75}}{\sqrt{\alpha_f^2 v_3^2 (v^\Sigma_{75})^2+M_\eta^2}} .
\end{equation}
We mentioned already that some sectors have no $\eta$-mixing, with $c_L=c_{d^c}=0$. Furthermore the mixing only involves $f_{3}$ so $f_{l_{1,2}} \equiv f_{1,2}$.

With the $\eta$-mixing establishing the light-states, we considering for now just the terms
$\Psi_{\eta} \, \varphi \, \Psi_{\eta} + \frac{1}{\Lambda} (\phi_3 \Psi)  \varphi'  \Psi_\eta \, $ in the Lagrangian (\eq{lag}).
The desired  vacuum configuration for the Higgs multiplets  $\varphi$, $\varphi'$ is
$\vev{\varphi}=v_{5,\bar{5}}^\varphi$, $\vev{\varphi'} =v_{\bar{5}}^{\varphi'}$ (i.e. $\varphi'$ has no $5$ VEV, just $\bar{5}$). Going to the basis defined by \eq{basisquark} it is easy to see that the term $\Psi_{\eta} \, \varphi \, \Psi_{\eta}$ gives rise only to the following light-state mass term
\beq c_Q c_{u^c}\, v^\varphi_5\, \,u_{l_3} u^c_{l_3} ,
\eeq
that we identify with the top quark. The Clebsch-Gordan $\alpha_f$ are such that $c_Q= -c_{u^c}$ \cite{Nath:2003rc}.
On the other hand the term $\frac{1}{\Lambda} (\phi_3 \Psi)  \varphi'  \Psi_\eta $   gives rise  to two light-state mass terms
\beq c_Q \, v_3\,v^{\varphi'}_{\bar{5}}\, \,d_{l_3} d^c_{l_3}+  c_{e^c} \, \, v_3\,v^{\varphi'}_{\bar{5}}\, \,e_{l_3} e^c_{l_3} ,
\eeq
which we identify as the bottom and the tau respectively. Bottom and tau unification and the  hierarchy between $m_t$ and $m_b$ is realized with
\bea
c_Q \simeq  -c_{e^c} & \simeq & 1 , \nn\\
v_3 v^{\varphi'}_{\bar{5}} / v^\varphi_5 &\simeq &1/10  ,
\eea
which requires that 
\bea
\label{defangap}
M_\eta& \ll& v_{3} v_{75}^\Sigma , \nn\\
s_Q &=& s_{u^c}\simeq 3 \sqrt{2} \frac{M_\eta}{v_{3} v_{75}^\Sigma} , \nn\\
s_{e^c} &\simeq & \sqrt{2} \frac{M_\eta}{v_{3} v_{75}^\Sigma} ,
\eea
having made explicit the Clebsch-Gordan coefficients $\alpha_f$. It is convenient to define $r \equiv \frac{M_\eta}{v_{3} v_{75}^\Sigma}$.
Note that a term of the kind  $(\phi_{A} \Psi) \varphi' \Psi_\eta $  would not change the top quark mass term even if $\vev{\varphi'}$ had a $5$ VEV - the contribution would vanish as it is proportional to $(c_Q+ c_{u^c})=0$.

The remaining Yukawa terms contained in \eq{lag}
\beq
\frac{1}{\Lambda^{2}}(\Psi\,\phi_{23})\, \rho\,(\Psi \,\phi_{23} )+\frac{1}{\Lambda^{2}}(\Psi\,\phi_O)\,\tilde{\varphi}\, (\Psi\,\phi_O)+\frac{1}{\Lambda^{3}}(\Psi\,\phi_O) \rho (\Psi\,\phi_O) \phi_{0} ,
\eeq
contribute mass terms to the lighter generations. Both terms with $\phi_{O}$ are similar in structure and can be considered together ($x_{f}$ in the following matrix - the distinct Higgs leads to family specific factors that are different for each family as we see in eq.(\ref{epsf})).
With the  familon vacuum configuration $\frac{\vev{\phi_{23}}}{\Lambda} \equiv \alpha (0,1,-1)$, $\frac{\vev{\phi_O}}{\Lambda} \equiv \epsilon  (2,-1,-1)$ and $\frac{\vev{\phi_{0}}}{\Lambda} \equiv v_{0}$, the Dirac mass matrices  present the general form

\begin{eqnarray}
\label{quark2}
M^f_{LR} & =&  \left( \begin{array}{ccc} 4 x_f  &- 2 x_f &- 2 s_{f^c} x_f  \\ -2 x_f &y_f+x_f &s_{f^c}(- y_f+x_f )\\ -2 s_F \, x_f  &s_F(-y_f+x_f )&s_F s_{f^c}(y_f+x_f )+z_f \end{array}\right) .
\end{eqnarray}
$x_{f}$ encodes the $\phi_{O}$ contributions, $y_{f}$ the $\phi_{23}$ contributions, and $z_{f}$ the leading order contribution to the third generation that was already discussed in detail. The desired Higgs VEV configuration is $v^{\tilde{\varphi}}_{5}$, $v^{\tilde{\varphi}}_{\bar{5}}$ for $\tilde{\varphi}$, while $\rho$ develops $v^{\rho}_{\bar{45}}$, $v^{\rho}_{5}$ in addition to the singlet ($v^{\rho}_{1}$ as previously seen in section \ref{neutrinos}).
More specifically, for each charged fermion family we have
\begin{eqnarray}
\label{quark2esp}
M^u_{LR} &=& \left( \begin{array}{ccc}4 \epsilon_u^2 v_5^{\tilde{\varphi}} &-2 \epsilon_u^2 v_5^{\tilde{\varphi}}&  -  6 \sqrt{2} r  \epsilon_u^2 v_5^{\tilde{\varphi}}\\  -2 \epsilon_u^2 v_5^{\tilde{\varphi}}& \alpha^2 v_5^\rho +\epsilon_u^2 v_5^{\tilde{\varphi}}& 3 \sqrt{2}r ( -\alpha^2 v_5^\rho + \epsilon_u^2 v_5^{\tilde{\varphi}})\\ -6 \sqrt{2}r \epsilon_u^2 v_5^{\tilde{\varphi}} &3 \sqrt{2} r ( -\alpha^2 v_5^\rho + \epsilon_u^2 v_5^{\tilde{\varphi}}) &18 r^2 ( \alpha^2 v_5^\rho +\epsilon^2 v_5^{\tilde{\varphi}} )+v_5^\varphi \end{array} \right) , \nn\\
&&\nn\\
M^d_{LR} &=& \left( \begin{array}{ccc}4 \epsilon_d^2 v_{\bar{5}}^{\tilde{\varphi}} &-2 \epsilon_d^2 v_{\bar{5}}^{\tilde{\varphi}}&  -  2   \epsilon_d^2 v_{\bar{5}}^{\tilde{\varphi}}\\  -2 \epsilon_d^2 v_{\bar{5}}^{\tilde{\varphi}}& \alpha^2 v_{\bar{45}}^\rho+ \epsilon_d^2 v_{\bar{5}}^{\tilde{\varphi}}&  ( -\alpha^2 v_{\bar{45}}^\rho + \epsilon_d^2 v_{\bar{5}}^{\tilde{\varphi}})\\ -6 \sqrt{2}r  \epsilon_d^2 v_{\bar{5}}^{\tilde{\varphi}} &3 \sqrt{2}r ( -\alpha^2 v_{\bar{45}}^\rho + \epsilon_d^2 v_{\bar{5}}^{\tilde{\varphi}}) &3 \sqrt{2} r  ( \alpha^2 v_{\bar{45}}^\rho +\epsilon_d^2 v_{\bar{5}}^{\tilde{\varphi}} )+v_{\bar{5}}^{\varphi'} \end{array} \right) , \nn\\
&&\nn\\
M^l_{LR} &=& \left( \begin{array}{ccc}4 \epsilon_l^2 v_{\bar{5}}^{\tilde{\varphi}} &-2 \epsilon_l^2 v_{\bar{5}}^{\tilde{\varphi}}&  -  2 \sqrt{2} r \epsilon_l^2 v_{\bar{5}}^{\tilde{\varphi}}\\  -2 \epsilon_l^2 v_{\bar{5}}^{\tilde{\varphi}}& -3\alpha^2 v_{\bar{45}}^\rho+ \epsilon_l^2 v_{\bar{5}}^{\tilde{\varphi}}&  \sqrt{2} r ( -3\alpha^2 v_{\bar{45}}^\rho + \epsilon_l^2 v_{\bar{5}}^{\tilde{\varphi}})\\ -2   \epsilon_l^2 v_{\bar{5}}^{\tilde{\varphi}} & ( -3\alpha^2 v_{\bar{45}}^\rho + \epsilon_l^2 v_{\bar{5}}^{\tilde{\varphi}}) &\sqrt{2} r  (3  \alpha^2 v_{\bar{45}}^\rho +\epsilon_l^2 v_{\bar{5}}^{\tilde{\varphi}} )+v_{\bar{5}}^{\varphi'} \end{array} \right) ,
\end{eqnarray}
noting that $r \equiv \frac{M_\eta}{v_{3} v_{75}^\Sigma}$ appears along with the appropriate Clebsch-Gordan factors through $s_{f^c}, s_F$ (\eq{defangap}), due to $\eta$-mixing. We have absorbed the complex Yukawa parameters in the Higgs scalar VEVs. We have also defined the family specific
\bea
\label{epsf}
\epsilon^2_u&=& \epsilon^2 (1+ v_0 v_5^\rho/v_5^{\tilde{\varphi}}) , \nn\\
\epsilon^2_d&=& \epsilon^2 (1+ v_0 v_{\bar{45}}^\rho/v_{\bar{5}}^{\tilde{\varphi}}) , \nn\\
\epsilon^2_l&=& \epsilon^2 (1-3 v_0 v_{\bar{45}}^\rho/v_{\bar{5}}^{\tilde{\varphi}}) ,
\eea
to condense the two distinct (but similar) $\phi_{O}$ contributions encoded in $x_{f}$.

The three charged fermion mass matrices of \eq{quark2esp} are diagonalized by
\beq
U_L^{f\dag} \, M_{LR}^f U_R^f=  Diag (m^f_1,m^f_2,m^f_3) ,
\eeq
where  $U_L^{u,d}$ give us the CKM mixing matrix  in the quark sector defined as $V_{CKM}= U_L^{u\dag} U_L^d$ while $U_L^l$ produces corrections to TBM mixing in the lepton sector.
In order to  recover the typical FN textures for the charged fermion  we need $\epsilon v_{5,\bar{5}}^{\tilde{\varphi}}< \alpha v_{5,\bar{45}}^\rho<v_5^\varphi, v_{\bar{5}}^{\varphi'}$, so we can use $Det (M_{LR}^{f} M_{LR}^{f \dag}) = (m_1^{f}  m_2^{f}  m_3^{f})^{2} $  to get approximated  expressions for the charged fermion masses
\bea
\label{masses}
(m_u,m_c,m_t)&\simeq& (4 \epsilon_u^2 v_5^{\tilde{\varphi}} , \alpha^2  v^\rho_5,v_5^\varphi) , \nn\\
(m_d,m_s,m_b)&\simeq&( 4 \epsilon_d^2 v_{\bar{5}}^{\tilde{\varphi}} , \alpha^2  v^\rho_{\bar{45}},v_{\bar{5}}^{\varphi'}) , \nn\\
(m_e,m_\mu,m_\tau)&\simeq&( 4 \epsilon_l^2 v_{\bar{5}}^{\tilde{\varphi}} , 3 \alpha^2 v^\rho_{\bar{45}},v_{\bar{5}}^{\varphi'}) .
\eea
Eq.(\ref{masses}) correctly gives bottom-tau unification and $m_\mu \simeq 3 m_s$.  Moreover in order to have  $ m_\mu/m_\tau \sim \lambda^2$ we need $\alpha\sim \lambda $ for $v^\rho_{\bar{45}} \sim v^{\varphi'}_{\bar{5}}$, which in turn requires  $v_5^\rho\sim\lambda^2 v_5^\varphi$ to recover $m_c/m_t \sim \lambda^4$.

The LH mixing matrices are approximately given by 
\bea
\label{massmix}
U_L^{u\dag} &\simeq& \left(\begin{array}{ccc} 1&\mathcal{O}(\epsilon_u^2  v_5^{\tilde{\varphi}}/(\alpha^2v_5^\rho ))  & \mathcal{O}( 10 \epsilon_u^2 r v_5^{\tilde{\varphi}}/v_5^\varphi ) \\
-\mathcal{O}(\epsilon_u^2 v_5^{\tilde{\varphi}}/(\alpha^2 v_5^\rho)) & 1& \mathcal{O}(\alpha^2 v_5^\rho/v_5^\varphi) \\
-\mathcal{O}( 10 \epsilon_u^2 r v_5^{\tilde{\varphi}}/v_5^\varphi ) &- \mathcal{O}(\alpha^2 v_5^\rho/v_5^\varphi)& 1  \end{array} \right) , \nn\\
U_L^{d\dag} &\simeq& \left(\begin{array}{ccc}  1&\mathcal{O}(10 \epsilon_d^4  v_{\bar{5}}^{\tilde{\varphi}} /  (\alpha^4 v_{\bar{45}}^{\rho}))&\mathcal{O}(\epsilon_d^2 v_{\bar{5}}^{\tilde{\varphi}}/ v_{\bar{5}}^{{\varphi}}) \\
- \mathcal{O}(10 \epsilon_d^4  v_{\bar{5}}^{\tilde{\varphi}} / ( \alpha^4 v_{\bar{45}}^{\rho}))
 & 1& \mathcal{O}(\alpha^2 v_{\bar{45}}^\rho / v_{\bar{5}}^{\varphi'})\\
-\mathcal{O}(\epsilon_d^2 v_{\bar{5}}^{\tilde{\varphi}}/ v_{\bar{5}}^{{\varphi}}) &- \mathcal{O}(\alpha^2 v_{\bar{45}}^\rho/ v_{\bar{5}}^{\varphi'})& 1 \end{array} \right) , \nn\\
U_L^{l\dag} &\simeq& \left(\begin{array}{ccc} 1& \mathcal{O}( \epsilon_l^2  v_{\bar{5}}^{\tilde{\varphi}} /  (3 \alpha^2 v_{\bar{45}}^{\rho})) & \mathcal{O}(10 r \epsilon_l^2 v_{\bar{5}}^{\tilde{\varphi}}/ v_{\bar{5}}^{{\varphi'}})  \\
-\mathcal{O}( \epsilon_l^2  v_{\bar{5}}^{\tilde{\varphi}} / (3 \alpha^2 v_{\bar{45}}^{\rho})) & 1& \mathcal{O}(3 \alpha^2 v_{\bar{45}}^\rho/ v_{\bar{5}}^{\varphi'})\\
- \mathcal{O}(10 r \epsilon_l^2 v_{\bar{5}}^{\tilde{\varphi}}/ v_{\bar{5}}^{{\varphi}}) &- \mathcal{O}( 3 \alpha^2 v_{\bar{45}}^\rho/ v_{\bar{5}}^{\varphi'})& 1 \end{array} \right) .
\eea
Note that $U_L^d$ and $U_L^l$ have rather different $12$ entries - the orthogonality between $\vev{\phi_{23}}$ and $\vev{\phi_O}$ cancels the contribution proportional to $\alpha^2 \epsilon^2$ in the entry $12$ of $M_{LR}^d M_{LR}^{d \dagger}$, but the $\eta$-mixing of the third family of the RH leptons enables a $\alpha^2 \epsilon^2$ contribution in the $12$ of in $M_{LR}^l M_{LR}^{l \dagger}$.

With $v^\rho_{\bar{45}} \sim v^{\varphi'}_{\bar{5}}$ (previously chosen when fitting $m_{\mu}/m_{\tau}$) we automatically get  from $M^d_{LR}$ the correct magnitude for $\theta_{23}^d$ in eq.(\ref{massmix})
\beq 
\theta_{23}^d \sim \theta_{23}^{CKM} \sim \lambda^2\,.
\eeq

In order to fit the light family masses and the Cabibbo angle it is necessary that the latter arises  from $U_L^d$. Remembering that $v_0=\vev{\phi_0}/\Lambda \sim \lambda^3$ from the neutrino sector,  we need $v_5^{\tilde{\varphi}}\sim \lambda^2 v_5^\rho$, $v_{\bar{5}}^{\tilde{\varphi}}\sim(\lambda^3-\lambda^2)  v_{\bar{45}}^\rho$ and $\epsilon_u\sim\epsilon_d\sim \epsilon_l\sim \lambda$. Moreover, since $r \ll 1$,  even  $\theta_{13}^{CKM}$ arises mainly by $U_L^d$. By substituting  the values indicated  into the expressions for $U_{L}^{u,d,l}$ given   in \eq{massmix} we get\bea
\label{app}
\theta_{12}^{CKM}\sim \lambda& \quad , \quad &
\theta_{13}^{CKM}\sim \mathcal{O}(\lambda^5-\lambda^4) , \nn\\
\theta_{12}^l \sim   \mathcal{O}(\lambda^2/3)&\quad , \quad&
\theta_{23}^l \sim   \mathcal{O}(\lambda^2)\quad , \quad \theta_{13}^l <   \mathcal{O}(\lambda^5) , \nn\\
\frac{m_u}{m_t}\sim \mathcal{O}(\lambda^6) &\quad , \quad&
\frac{m_d}{m_b}\sim \mathcal{O}(\lambda^4) ,
\eea
where $\theta_{12,23,13}^l$ are   the deviations  of $U_L^{l \dag}$ from the identity.  From \eq{app}  we can estimate the amount of shifting of the lepton mixing  from exact TBM mixing. At order $\mathcal{O}(\lambda^2)$ we get

\beq
U_{lep} =\left( \begin{array}{ccc} \sqrt{\frac{2}{3}} -\frac{\lambda^2}{3\sqrt{6}} &\frac{1}{\sqrt{3}}+\frac{\lambda^2}{3\sqrt{3}} &-\frac{\lambda^2}{3\sqrt{2}} \\-\frac{1}{\sqrt{6}} -(\sqrt{\frac{2}{27}}+\sqrt{\frac{3}{2}})\lambda^2& \frac{1}{\sqrt{3}} - \frac{8\sqrt{3}}{9} \lambda^2 &-\frac{1}{\sqrt{2}}+\frac{3\sqrt{2}}{2} \lambda^2\\ -\frac{1}{\sqrt{6}} +\sqrt{\frac{3}{2}}\lambda^2 &\frac{1}{\sqrt{3}} -\sqrt{3}\lambda^2 &\frac{1}{\sqrt{2}} +\frac{3 \sqrt{2}}{2}\lambda^2 \end{array}\right) ,
\eeq
that gives
\bea
\sin\theta_{12}^2 &=& \frac{1}{3}+\frac{2}{9} \lambda^2 + \mathcal{O}(\lambda^4) , \nn\\
\sin\theta_{23}^2 &=& \frac{1}{2}-3 \lambda^2 + \mathcal{O}(\lambda^4) ,\nn\\
\sin\theta_{13}^2 &=& \mathcal{O}(\lambda^4) .
\eea
The comparison between  the analytical expressions  we get with  the neutrino fit data \cite{Maltoni:2004ei} shows that  we are inside the   $2-\sigma$ range for all the three angles.

Finally, the  degeneracy between the down quark and  the electron mass  is solved  by  having   $\epsilon^l\neq \epsilon^d$ and therefore $m_e$ can be correctly fitted.

\section{Vacuum alignment \label{VEVA}}


In the previous sections we assumed that $SO(10)$ is broken directly to $SU(3)\times SU(2)\times U(1)$ through the vev of the $75$ of $SU(5)$ contained in $\Sigma$ and the VEVs of the SM singlets contained in $\xi$ and $\rho$. In addition we assumed that $\vev{\Sigma}>> \vev{\xi},\vev{\rho}$  to recover the correct neutrino mass matrix and the  absolute neutrino mass scale.  The construction of the superpotential that reaches the correct breaking pattern goes beyond the purpose of this work. However it can be obtained using established  strategies already  applied in the literature \cite{Rajpoot:1980xy,Barr:1981qv,Chang:1985zq,Bajc:2004xe,Goh:2004fy,Babu:2005gx}. 
Since we break $SO(10)$ directly to the SM the GUT scale of the model coincides with the $SU(5)$ one, that is $M_{GUT} \sim 2 \times 10^{16}$ GeV. However the presence of the $210$ with respect to the minimal $SO(10)$ GUT model \cite{Bajc:2004xe} and the requirement of preserving  the gauge coupling  pertubativity forces the model cut-off scale $\Lambda$ to  be approximately $10^{17}$, few orders below the usual one. In principle familons and Higgs scalars could have different cut-off scales, $\Lambda_G$ and $\Lambda_F$ respectively, but we assume that they coincide ($\Lambda_G= \Lambda_F=\Lambda$).

\subsection{Familon alignment \label{f_align}}

The pattern of VEVs displayed in eq.(\ref{f_vevs}) plays a crucial role in our model, and we now discuss how to obtain it. A relatively simple way to obtain the desired relies on the use of a discrete
non-Abelian subgroup of $SU(3)_{F}$ (alternatively, in $SU(3)_{F}$ it is possible to obtain the pattern by adding several alignment fields, as in \cite{Ivo1}).
The alignment mechanism we use is based on $\Delta(27)$, belonging to the $\Delta (3 n^2)$ family of groups \cite{Luhn:2007uq}, and the method proposed here is rather similar to the one originally presented in \cite{Ivo3}.
Higher-order invariant terms can arise in the scalar potential through SUSY breaking soft terms and break the degeneracy of VEVs that would exist in the continuous group - these invariants are allowed by the discrete FS (but not by $SU(3)_{F}$). These terms are very small but are the only terms that distinguish VEV directions and so must be considered in the alignment discussion. On the other hand, the Yukawa superpotential is approximately invariant under $SU(3)_{F}$: the higher-order terms can be neglected compared to the terms allowed by the continuous FS, such that the Lagrangian is given by eq.(\ref{lag}) to good approximation.

As discussed in section \ref{mod}, some of the familons acquire VEVs with larger magnitudes (namely $\phi_{123}$, but also $\phi_{3}$). The leading $D-$terms for these familons leads to a potential
\begin{equation}
V\left( \phi_{A} \right) =\alpha_{A} m^{2}\sum_{i}\left\vert \phi_{A} ^{i}\right\vert
^{2} + \beta_{A} m^{2}\left\vert \sum_{i}\left\vert \phi_{A} ^{i}\right\vert
^{2}\right\vert ^{2}
+\gamma_{A} m^{2}\sum_{i}\left\vert \phi_{A} ^{i}\right\vert^{4} ,
\label{potential}
\end{equation}%
where $m$ is the gravitino mass. These are soft terms that arise only if SUSY is broken (which is why $m^{2}$ appears on every term).
The coefficient $\alpha_{A}$ is radiatively driven negative near the scale $\Lambda$, triggering a VEV for $\phi_{A}$.
The second term is generated at one-loop order if the superpotential contains a term of the
form $Y \Xi \sum_{i}\phi_{A}^{i} \chi_{i}$ where Y is a FS singlet, with $\chi_{i}$ (charged under the FS) and $\Xi$ being massive chiral superfields (that go in the loop). The two first terms in eq.(\ref{potential}) are invariant under the continuous group $SU(3)_{F}$ and, with $\alpha_{A}$ negative, generate $\vev{\phi_{A}}$ with a constant non-zero magnitude $x$ of the order of $\Lambda$.
The third term breaks $SU(3)_{F}$ but is consistent with $\Delta(27)$. It will be generated if the underlying theory contains a superpotential term of the form $Z \sum_{i} \phi_{A}^{i} \varphi^{i} \varphi^{i}$, where $Z$ is a singlet of $\Delta(27)$ and $\varphi^{i}$ is a massive chiral superfield (that goes in the loop) with the appropriate FS assignments. The
resulting third term of eq.(\ref{potential}) splits the vacuum degeneracy.
The minimum for $\gamma_{A}$ positive has $\left\vert
\langle \phi_{A}^{i} \rangle \right\vert =x(1,1,1)/\sqrt{3}$ while for $\gamma $ negative
$\left\vert \langle \phi_{A}^{i} \rangle \right\vert =x(0,0,1)$ (the non-zero entry defines the preferred direction).
The phases are unspecified as these terms do not establish any preferred phase.

This provides a mechanism to generate the vacuum alignment
of $\phi _{3}$ and $\phi_{123}$ as each will have a potential of the form
in eq.(\ref{potential}), provided they acquire large VEVs. The structure of eq.(\ref{f_vevs}) results if 
$\gamma _{3}$ is positive and $\gamma _{123}$ is negative (and by definition $\vev{\phi _{3}}$ lies in the third direction). In order for the correct alignment to be reached, the terms featuring just the respective familon need to dominate over similar quartic terms mixing separate familons which may be present (e.g. $\phi_{123}^{i} \phi_{3_{i}}^{\dagger} \phi_{3}^{j} \phi_{123_{j}}^{\dagger}$ or $\phi_{23}^{i} \phi_{3_{i}}^{\dagger} \phi_{3}^{j} \phi_{23_{j}}^{\dagger}$). For this reason the magnitudes of $\vev{\phi_{123}}$ and $\vev{\phi_{3}}$ are required by naturalness to be somewhat larger than $\vev{\phi_{23}}$ and $\vev{\phi_{O}}$, which arise at a scale slightly smaller than $\Lambda$.

For $\phi_{23}$ to receive the correct alignment, we need to introduce an additional familon $\phi_{1}$ which receives a large VEV of order $\Lambda$ (just like $\phi_{123}$ and $\phi_{3}$), with positive $\gamma_{1}$ and taking a direction which we define to be the first - $\left\vert \langle \phi_1 ^{i} \rangle \right\vert = x(1,0,0)$ (to justify why $\vev{\phi_{1}}$ and $\vev{\phi_{3}}$ have distinct directions, the mixed quartic terms involving $\phi_{3}$ and $\phi_1$ must favor their VEVs to be orthogonal by having a positive coefficient).
The terms responsible for aligning $\phi_{23}$ in the desired direction arise just like the $\beta$ quartics of eq.(\ref{potential}), but are naturally dominant over the unmixed $\beta_{23}$ term: the dominant terms must be $\beta'_{1} m^{2} \phi_{1}^{i} \phi_{23_{i}}^{\dagger} \phi_{23}^{j} \phi_{1_{j}}^{\dagger}$ and $\beta'_{123} m^{2} \phi_{123}^{i} \phi_{23_{i}}^{\dagger} \phi_{23}^{j} \phi_{123_{j}}^{\dagger}$. A positive $\beta'_{1}$ term favors $\vev{\phi_{23}^{1}}=0$. A positive $\beta'_{123}$ term leads to the orthogonality of VEVs of $\phi_{123}$ and $\phi_{23}^{\dagger}$.

We introduce also an alignment field $X$. Due to the symmetry content, the only superpotential term directly relevant to our alignment purposes is $X \sum_{i} \phi_{23}^{i} \phi_{23}^{i} \phi_{23}^{i}$ ($X$ has $U(1)_{F}$ of $-3 q_{23}$). This invariant is allowed by $\Delta(27)$ and the corresponding $F-$term produces the vacuum condition $\sum_{i} (\phi_{23}^{i})^{3}=0$.
This condition is only satisfied for specific relative phases of the $\vev{\phi_{23}}$ components - the cube of the entries must close a triangle in the complex plane. With the soft terms favoring a non-vanishing VEV with $\vev{\phi_{23}^{1}}=0$, it is in fact a degenerate triangle and we conclude that one of the discrete set of possible solutions is $\vev{\phi_{23}^{2}}=-\vev{\phi_{23}^{3}}$. In this case the correct orthogonality condition is obtained from $\beta'_{123}$, fixing the relative phases, with only the global phases remaining unknown.

Finally $\phi_{O}$ is aligned correctly if the dominant terms governing its alignment are $\beta''_{1} m^{2} \phi_{1}^{i} \phi_{O_{i}}^{\dagger} \phi_{O}^{j} \phi_{1_{j}}^{\dagger}$ with negative $\beta''_{1}$, and $\beta''_{123} m^{2} \phi_{123}^{i} \phi_{O_{i}}^{\dagger} \phi_{O}^{j} \phi_{123_{j}}^{\dagger}$ with positive $\beta''_{123}$.

\section{Conclusion}

In this paper we studied some of the possibilities provided by considering an extended seesaw scenario in a family symmetry grand unified model, presenting a specific case with phenomenologically viable fermion masses and mixings.

Neutrino mixing is tri-bi-maximal from the combination of a specific realization of the extended seesaw mechanism with a specific vacuum alignment configuration (directly related to the structure of the discrete non-Abelian family symmetry used).
The charged lepton mixing angles are small and produce slight deviations from tri-bi-maximal mixing.

The charged fermion mass terms produce a structure that can fit the mass hierarchies and the CKM mixing angles (consistently with preserving near tri-bi-maximal leptonic mixing, as described above).

The model is fairly complicated, with a large field content. However it demonstrates the potential benefits of considering extended seesaw realizations in this class of unified models with a family symmetry. In the model presented, the Higgs content is relatively less constrained: the phenomenologically required separation of the neutrino sector from the charged fermions - despite their unification in the same multiplet - is relatively easy to achieve by using a slightly enlarged matter content.

\section*{Acknowledgments}
The work of FB has been partially supported by MEC-Valencia  MEC grant FPA2008-00319/FPA, by European Commission Contracts
MRTN-CT-2004-503369 and ILIAS/N6 RII3-CT-2004-506222 and by the foundation for Fundamental Research of Matter (FOM) and the National Organization for Scientific Research (NWO).
The work of IdMV was supported by FCT under the grant SFRH/BPD/35919/2007.
The work of IdMV was partially supported by FCT through the projects
POCI/81919/2007, CERN/FP/83503/2008
and CFTP-FCT UNIT 777  which are partially funded through POCTI
(FEDER) and by the Marie Curie RTN MRTN-CT-2006-035505.
IdMV would like to thank IFIC-UV (Valencia) for their warm hospitality and
support during a short stay when this project was initiated. 

\bibliography{refs}
\bibliographystyle{hepv}

\end{document}